\documentclass[prl,twocolumn,showpacs,amsmath,amssymb,preprintnumbers]{revtex4}
\input epsf
\usepackage{epsfig,psfrag}

\usepackage{graphicx}

\usepackage{color}
\usepackage{ulem}

\newcommand{\eps}{\epsilon}

\def\be{\begin{equation}}
\def\ee{\end{equation}}
\def\bea{\begin{eqnarray}}
\def\eea{\end{eqnarray}}
\def\alphaeff{a}

\begin{document}
\title{The three-loop cusp anomalous dimension in QCD}

\author{Andrey Grozin}
\affiliation{
     Budker Institute of Nuclear Physics SB RAS, Novosibirsk, Russia\\
        Novosibirsk State University, Novosibirsk, Russia}
        \email{A.G.Grozin@inp.nsk.su}
\author{Johannes M. Henn}
\affiliation{Institute for Advanced Study, Princeton, NJ 08540, USA}
\email{jmhenn@ias.edu}
\author{Gregory P. Korchemsky}
\affiliation{Institut de Physique Th\'eorique, CEA Saclay, 91191 Gif-sur-Yvette Cedex, France}
\email{Gregory.Korchemsky@cea.fr}
\author{Peter Marquard}
\affiliation{Deutsches Elektronen-Synchrotron, DESY, Platanenallee 6, D15738 Zeuthen, Germany}
\email{peter.marquard@desy.de}

\preprint{IPhT-T14-111, DESY 14-148, SFB/CPP-14-64, LPN14-104.}

\date{\today}

\begin{abstract}
  We present the full analytic result for the three-loop
  angle-dependent cusp anomalous dimension in QCD. With this result,
  infrared divergences of planar scattering processes with massive
  particles can be predicted to that order.  Moreover, we define a
  closely related  quantity in terms of an
  effective coupling defined by the light-like cusp anomalous
    dimension. We find evidence that this quantity is universal for
  any gauge theory, and use this observation to predict the non-planar
  $n_{f}$-dependent terms of the four-loop cusp anomalous dimension.
\end{abstract}


\maketitle


{The cusp anomalous dimension is an ubiquitous quantity in four-dimensional gauge theories. It governs
the dependence of a cusped Wilson loop on the ultraviolet cut-off  \cite{PolyakovCusp} 
and appears in many physical quantities.  
In particular, it controls the infrared asymptotics of 
scattering amplitudes and form factors involving massive particles \cite{Korchemsky:1991zp,hqetreviews}, see e.g. \cite{Czakon:2009zw} for a recent application to top quark pair production.
The two-loop result for this fundamental quantity has been known for more than 25 years 
\cite{twoloop}.
Here we report on the full result for the cusp anomalous dimension in QCD at three 
loops.}

\begin{figure}
 \centerline{\psfrag{v1}[cc][cc]{$v_1$} \psfrag{v2}[cc][cc]{$v_2$} 
 \psfrag{phi}[cc][cc]{$\phi$}
  \includegraphics[width = 50mm]{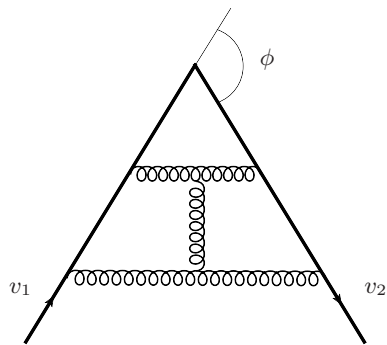}
 }
  \caption{Sample Feynman diagram producing a contribution to 
the three-loop cusp anomalous dimension in QCD. Thick
  lines denote two semi-infinite segments forming a cusp of angle 
$\phi$, and wavy
  lines represent gauge fields.}
  \label{figurenf}
\end{figure}

To compute the cusp anomalous dimension,
we consider the vacuum expectation value of the Wilson line operator 
\begin{align}\label{W-bosonic}
W ={1\over N}\langle 0|\,{\rm tr} \bigg[ P \exp\left(i \oint_C dx\cdot A(x)   \right)\bigg] |0\rangle \,,
\end{align}
with $A_\mu(x) = A_\mu^a (x) T^a$ and $T^a$ being the generators of the fundamental representation
of the $SU(N)$ gauge group.
Here the integration contour $C$ is formed by two segments along space-like directions 
$v_{1}^\mu$ and $v_{2}^\mu$ (with $v_1^2= v_2^2 =1$), with (Euclidean space) cusp angle $\phi$,
\begin{equation}
{\cos \phi = v_{1} \cdot v_{2}} \,,
\end{equation}
cf. Fig.~\ref{figurenf}. Perturbative corrections to the Wilson loop (\ref{W-bosonic}) contain both ultraviolet (cusp) and 
infrared divergences. We employ dimensional regularization with $D=4-2\epsilon$ to regularize the former
and introduce an infrared cut-off using the heavy quark effective theory (HQET) framework. The cusp anomalous
dimension $\Gamma_{\rm cusp}$ is extracted as the residue at the simple pole $1/\epsilon$ in the corresponding renormalization factor.

It depends on the cusp angle 
$\phi$, the strong coupling constant $\alpha_{s} = g_{\rm YM}^2/(4 \pi)$,
and on $SU(N)$ color factors. It is convenient to introduce the complex variable 
\begin{align}\label{defy}
x=e^{i \phi}\,, \qquad 2 \cos \phi =  x+{1}/{x} \,.
\end{align}  
In Euclidean space $|x|=1$, whereas for Minkowskian angles $\phi=i\theta$ (with $\theta$ real) the variable $x$ can take arbitrary nonnegative values. 
Due to the symmetry $x\to 1/x$ of the definition (\ref{defy}), we can assume $0<x<1$ without loss of generality.

We chose to perform the calculation in momentum space. We  
generated all Feynman diagrams contributing to $W$ up to three loops, in an arbitrary covariant gauge.
This was done with the help of the computer programs QGRAF and FORM \cite{Nogueira:1991ex}.
Using integration by parts relations \cite{IBP},
we found that a total of $71$ master integrals was required. We derived differential equations
for them in the complex variable $x$ defined in (\ref{defy}).
Switching to a basis of master integrals $\vec{f}(x,\epsilon)$ as suggested in ref. \cite{Henn:2013pwa},
we found the expected canonical form of the differential equations,
\begin{align}\label{DEhqet}
\partial_{x} \, \vec{f}(x,\eps) =  \eps \, \left[ \frac{a}{x} + \frac{b}{x+1} + \frac{c}{x-1} \right] \vec{f}(x,\eps)\,,
\end{align}
with constant ($x-$ and $\eps-$independent) matrices $a,b,c$.

Eq. (\ref{DEhqet}) has four regular singular points in $x$, namely $0,1,-1,$ and $\infty$. 
Thanks to the $x \to {1}/{x}$ symmetry of the definition (\ref{defy}), only the first
three are independent. They correspond, in turn, to the light-like limit (infinite {Minkowski} angle), to the zero angle limit, and to the antiparallel lines limit.
Requiring that the integrals be nonsingular in the straight-line case $x=1$ allowed us to fix all except one boundary conditions, and we obtained the remaining one from ref. \cite{Chetyrkin:2003vi}. 

It follows from (\ref{DEhqet}) that the solution for $\vec{f}$ in the $\eps-$expansion can be written in terms of iterated integrals with integration kernels $dx/x, dx/(x-1), dx/(x+1)$. The latter integrals are known as harmonic polylogarithms $H_{n_1 \dots n_k} (x)$ \cite{Remiddi:1999ew}.
The indices $n_{i}$ can take values $0,1,-1$, corresponding to the three integration kernels, respectively.

To express our results up to three loops, we introduce the following functions
\footnote{The function $A_{5}$ (and $A_{1}$ and $A_{3}$) appeared previously in the result for the cusp anomalous dimension in $\mathcal{N} = 4$ super Yang-Mills (SYM) for a locally supersymmetric Wilson line operator \cite{3loopsN4SYM}.},
\begin{widetext}
\begin{equation}\label{A-B}
\begin{aligned}  
{A}_1(x) =&  \xi \,   \frac{1}{2} H_{1}(y)\,,\quad\quad
%
{A}_2(x)  = \, \left[ \frac{\pi^2}{3} + \frac{1}{2} H_{1,1}(y) \right]  + \xi \left[- H_{0,1}(y) -\frac{1}{2} H_{1,1}(y) \right] \,, \\
%
{A}_{3}(x) =& \;\;  \xi \, \left[ -\frac{\pi^2}{6} H_{1}(y) - \frac{1}{4} H_{1,1,1}(y)  \right]  
 +  \xi^2\, \left[ \frac{1}{2} H_{1,0,1}(y) + \frac{1}{4} H_{1,1,1}(y) \right]  \,,\\
%
%
{A}_{4}(x) =&
 \, \left[  -\frac{\pi^2}{6} H_{1,1}(y) -\frac{1}{4} H_{1,1,1,1}(y)  \right] +  \\
& + \xi \left[  \frac{\pi^2}{3} H_{0,1}(y) +\frac{\pi^2}{6} H_{1,1}(y) + 2 H_{1,1,0,1}(y) +\frac{3}{2} H_{0,1,1,1}(y)+ \frac{7}{4} H_{1,1,1,1}(y) + 3 \zeta_3 H_{1}(y)  \right]  \\
&+  \xi^2 \left[ -2 H_{1,0,0,1}(y)-2 H_{0,1,0,1}(y)-2 H_{1,1,0,1}(y) - H_{1,0,1,1}(y) - H_{0,1,1,1}(y) -\frac{3}{2} H_{1,1,1,1}(y) \right]\,,\\
{A}_{5}(x) =&\, \xi \left[  \frac{\pi^4}{12} H_{1}(y) + \frac{\pi^2}{4} H_{1,1,1}(y) +\frac{5}{8}  H_{1, 1, 1, 1, 1}(y) \right] +  \xi^2\left[ -\frac{\pi^2}{6} H_{1,0,1}(y) -\frac{\pi^2}{3} H_{0,1,1}(y) -\frac{\pi^2}{4}H_{1,1,1}(y)   \right.  \\
& \left. \;\;\;\; -H_{1,1,1,0,1}(y)  -\frac{3}{4} H_{1,0,1,1,1}(y) -H_{0,1,1,1,1}(y) -\frac{11}{8} H_{1,1,1,1,1}(y) -\frac{3}{2}\zeta_3 H_{1,1}(y)  \right] \\
& + \xi^3 \left[ H_{1,1,0,0,1}(y) + H_{1,0,1,0,1}(y) +H_{1,1,1,0,1}(y) + \frac{1}{2} H_{1,1,0,1,1}(y) +\frac{1}{2} H_{1,0,1,1,1}(y) +\frac{3}{4} H_{1,1,1,1,1}(y)  \right] \,,
 \\
B_{3}(x) =& \left[  - H_{1,0,1}(y) + \frac{1}{2} H_{0,1,1}(y) - \frac{1}{4} H_{1,1,1}(y)\right]  +  \xi \left[ 2 H_{0,0,1}(y) + H_{1,0,1}(y) + H_{0,1,1}(y) + \frac{1}{4} H_{1,1,1}(y) \right]\,,  \\
B_{5}(x) =& 
\frac{x}{1-x^2} \Big[-\frac{\pi^4}{60} H_{-1}(x) -\frac{\pi^4}{60} H_{1}(x) - 4H_{-1,0,-1,0,0}(x) + 4 H_{-1,0,1,0,0}(x) - 4 H_{1,0,-1,0,0}(x)    \\ &\quad\quad \quad\quad
+ 4 H_{1,0,1,0,0}(x) + 4 H_{-1,0,0,0,0}(x) + 4 H_{1,0,0,0,0}(x) + 2 \zeta_3  H_{-1,0}(x) + 2 \zeta_3 H_{1,0}(x)   \Big] \,, 
%
\end{aligned}
\end{equation}
where $\xi = (1+x^2)/(1-x^2)$ 
and $y=1-x^2$. 
The subscript of $A$ indicates the (transcendental) weight of the functions.
Moreover, we introduce the abbreviation $\tilde{A}_{i} = A_{i}(x) - A_{i}(1)$, and similarly for $\tilde{B}_{i}$. 

 \end{widetext}

Performing the three-loop computation, we reproduced the expected structure of UV divergences of $W$ in the $\overline{\rm MS}$ scheme, as well as the HQET wavefunction renormalization \cite{Chetyrkin:2003vi}, for arbitrary values of the gauge parameter in the covariant gauge. 
As yet another check, the dependence on the gauge parameter disappeared
for the cusp anomalous dimension.

Let us write the expansion in the coupling constant as 
\begin{align}
\Gamma_{\rm cusp}(\alpha_s , x) = 
\sum_{k\ge 1} \left(\frac{\alpha_s}{\pi}\right)^k \Gamma_{\rm cusp}^{(k)}(x)\,.
\end{align}
The previously known one- and two-loop \cite{twoloop} results can be written as
\begin{align}\label{result1loop}
\Gamma_{\rm cusp}^{(1)} ={}& C_{F} \,  \tilde{A}_1 \,, \\
\Gamma_{\rm cusp}^{(2)} ={}& 
\frac{1}{2}
 C_{F} C_{A} \left[ \tilde{A}_{3} + \tilde{A}_{2}  \right] \nonumber \\ & +  \left(\frac{67}{36} C_{F} C_{A} - \frac{5}{9} 
C_{F} T_{f} n_{f}  \right)  \tilde{A}_1 \,. \label{result2loop}
\end{align}
At three loops we find
\begin{equation}
\begin{aligned}\label{result3loop}
\Gamma_{\rm cusp}^{(3)} ={}& c_1\, C_{F} C_{A}^2 + c_{2}\, C_F (T_f n_f)^2 \\[1mm]
& + c_3\, C_{F}^2 T_f n_f + c_4\, C_{F} C_{A} T_{f} n_{f} \,,
\end{aligned}
\end{equation}
with
\begin{align} \notag  
  c_{1} ={}&{\frac{1}{4}} \left[ \tilde{A}_5  + \tilde{A}_{4} + \tilde{B}_{5} 
  +\tilde{B}_{3}  \right] \\
& +\frac{67}{36} \tilde{A}_{3}   +  \frac{29}{18}  \tilde{A}_{2} +  \left( \frac{245}{96} + \frac{11}{24} \zeta_3 \right) \tilde{A}_{1} \,,\\
c_{2} = {}& -\frac{1}{27}  \tilde{A}_1 \,, \quad
c_{3} = {}  \left( \zeta_3 - \frac{55}{48} \right)   \tilde{A}_1  \,, \label{resultc4} \\
c_{4} = {} & -\frac{5}{9}
 \left[ \tilde{A}_3 + \tilde{A}_{2} \right]  - \frac{1}{6} \left( 7 \zeta_{3} + \frac{209}{36} \right)  \tilde{A}_{1} 
\,. 
\end{align} 
Here $C_F=(N^2-1)/(2N)$ and $C_A=N$ are the quadratic Casimir operators of the $SU(N)$ gauge group in the fundamental and adjoint representation, respectively,  $n_f$ is the number of quark flavors, and $T_f=1/2$.

The following comments are in order. 
The cusp anomalous dimension has a branch cut for $x$ lying on the negative real axis.  
The results given in (\ref{result3loop}) are valid for $0< x <1$ and can be analytically continued
to other regions according to this choice of branch cuts \footnote{Note that $\Gamma_{\rm cusp}(\alpha_{s},-x)$ is expected to be related to $\Gamma_{\rm cusp}(\alpha_{s}, x)$ by crossing symmetry, i.e. the two functions should be equal, up to terms picked up by the analytic continuation. 
It turns out that all functions except $B_{5}$ contain rational factors that are symmetric under $x \to -x$, and therefore the harmonic polylogarithms in these expressions can be written with argument $1-x^2$, and positive indices only. In contrast, the factor $x/(1-x^2)$ contained in $B_{5}$ is antisymmetric under $x\to-x$, and so is the transcendental function multiplying it (up to terms coming from branch cuts).}.

The leading 
$n_f^2$
term in  (\ref{result3loop}) is in agreement with the known result \cite{Beneke:1995pq}.
We reported on the $n_f$-dependent part of (\ref{result3loop}) in \cite{Grozin:2014axa}.
The expression for the coefficient $c_1$ is new. 

As a check of our result, we can consider Minkowskian angles   and
take the light-like limit, $x=e^{-\theta}$ with $\theta\to\infty$, of eq.\,(\ref{result3loop}), where one expects the behavior \cite{limitx0}
\begin{align}\label{eqlimitx0}
 \Gamma_{\rm cusp}(\alpha_s ,x)  \stackrel{x \to 0}{=}{}  K(\alpha_s) \log (1/x) + {\mathcal{O}}(x^0) \,,
\end{align}
with  $K(\alpha_s)$ being  the light-like cusp anomalous dimension. To three loops, it is given by \cite{lightlike}
\begin{equation}\label{resultK}
\begin{aligned}
K^{(1)} =& \, C_{F} \,, \\
K^{(2)} =& \, C_A C_F \left( \frac{67}{36} - \frac{\pi^2}{12} \right) -\frac{5}{9} n_f T_f C_F \,,\\
K^{(3)} =& \,  
 C_{A}^2 C_{F}  \left( \frac{245}{96}- \frac{67 \pi^2}{216} + \frac{11 \pi^4}{720} + \frac{11}{24} \zeta_3 \right)   \\
& +  C_{A} C_{F} n_f T_{f} \left( -\frac{209}{216} + \frac{5 \pi^2}{54} - \frac{7}{6}\zeta_{3} \right)   \\
&+C^2_{F} n_f T_{f} \left( \zeta_3 -\frac{55}{48} \right) - \frac{1}{27} C_{F} (n_f T_f)^2  \,,
\end{aligned}
\end{equation}
where ${K(\alpha_s) = \sum_{m\ge 1} (\alpha_s/\pi)^m K^{(m)}}$.
We found perfect agreement for all terms.

Finally, in the antiparallel lines limit $\phi =  \pi-\delta\,, \delta \to 0$, one expects to recover the quark antiquark potential \cite{Kilian:1993nk} (at one
loop order lower compared to $\Gamma_{\rm cusp}$). Starting from eq. (\ref{result3loop}) we indeed find perfect agreement with the result quoted in the second ref. of \cite{Peter:1996ig}, up to terms proportional to the QCD $\beta-$function.

Our result for the cusp anomalous dimension is valid in the
$\overline{\rm MS}$ (dimensional regularisation) scheme. Going to 
  the $\overline{\rm DR}$ (dimensional reduction) scheme amounts to a
finite renormalisation of the coupling constant. We can introduce a
quantity $\Omega$ which is the same in both schemes by switching from
$\alpha_{s}$ to an ``effective coupling'' $a$,
\begin{equation}\label{defomega}
\begin{aligned}
 \Omega(\alphaeff ,x) :=   \Gamma_{\rm cusp}( \alpha_{s} ,x)\,,  \qquad \alphaeff  :=  \pi/C_{F} K(\alpha_{s}) \,,
\end{aligned}
\end{equation}
where $\Gamma_{\rm cusp}$ and $K(\alpha_{s})$ are evaluated in the same scheme (and for the same theory).
By construction, $\Omega$ has the universal limit
\begin{equation}\label{omegax0}
 \, \Omega( \alphaeff ,x)  \stackrel{x \to 0}{=}{}    \frac{\alphaeff }{\pi} C_{F}  \log (1/x) +{\mathcal{O}}(x^0) \,,
\end{equation}
as one can easily verify by comparing to eq. (\ref{eqlimitx0}).

Using the results up to three loops given in eqs. (\ref{result1loop}), (\ref{result2loop}), (\ref{result3loop}) and (\ref{resultK}), and expanding both sides of the first relation in (\ref{defomega}) to third order in $\alpha_{s}$, we find
\begin{align}
\Omega( \alphaeff, x) ={} & \, \frac{\alphaeff}{\pi} \, C_{F} \tilde{A}_1
+ \left( \frac{\alphaeff}{\pi}\right)^2 \, \frac{C_{A} C_{F} }{2}  \left[ \tilde{A}_3+ \tilde{A}_2 + \frac{\pi^2}{6} \tilde{A}_1 \right] \notag \\
& \hspace{-1.4cm} +  \left( \frac{\alphaeff}{\pi}\right)^3\, \frac{C_F C_A^2}{4}  \left[ \tilde{A}_5+ \tilde{A}_4 -\tilde{A}_2 + \tilde{B}_5 
+ \tilde{B}_3 
\right.  \label{result_omega} \\
& \hspace{-1.4cm} \quad\quad\quad\quad\quad\quad \;\;\;\; \left.  + \frac{\pi^2}{3} \tilde{A}_3 +  \frac{\pi^2}{3} \tilde{A}_2 - \frac{\pi^4}{180} \tilde{A}_1   \right]  + {\mathcal O}(\alphaeff^4) \,.\notag
\end{align}
Remarkably, this quantity is independent of $n_{f}$ to three loops!
Comparing to eq. (\ref{defomega}) we see that this means that e.g. all $n_{f}$ dependent terms in $\Gamma^{(3)}_{\rm cusp}$ are generated from lower-loop terms, when expanding 
$K(\alpha_s)$ in $\alpha_s$.

\begin{figure}
\psfrag{q}[cc][cc]{$\theta$}\psfrag{W}[cc][cc]{$\Omega( \alphaeff ,e^{-\theta})$}
  \includegraphics[width =80mm]{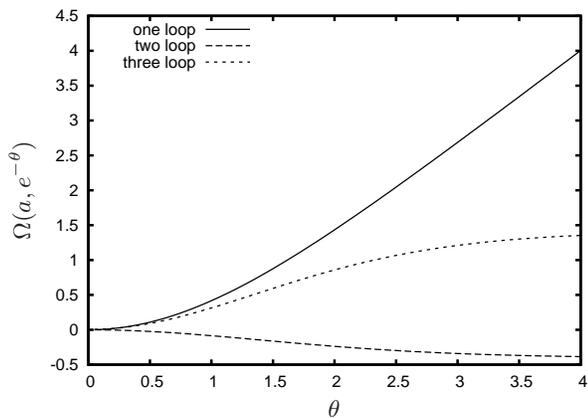}
  \caption{$\theta$ dependence of the cusp anomalous dimension $\Omega(\alphaeff,e^{-\theta})$ at 
  one (solid), two (dashed), and three (dotted) loops.
 }
  \label{figuretheta}
\end{figure}

In Fig.~\ref{figuretheta} we plot the one-, two- and three-loop coefficients of
$\Omega$ in an expansion of $\alphaeff/\pi$, 
for Minkowskian angles $\theta$, i.e. $x=e^{-\theta}$ 
for the range $\theta \in [0,4]$, and with the number of colors set to $N = 3$. 
Note that the $n_{f}$-dependence in QCD can be
obtained from eq. (\ref{defomega}), and amounts to a rescaling of the coupling.
At large $\theta$, the one-loop contribution displays the linear behavior of eq. (\ref{omegax0}), while the two- and three-loop contributions go to a constant, as expected.
In the small angle region, 
we have, 
\begin{align}
\Omega( \alphaeff ,e^{-\theta}) ={}& C_F\left[ 
\left( \frac{ \alphaeff}{\pi} \right) \frac{1}{3}  
 +  \left( \frac{ \alphaeff}{\pi} \right)^2 {C_A\over 4} \left( 1 - \frac{\pi^2}{9} \right)    \right.\\
& \left.  \hspace{-2.0cm} +  \left( \frac{ \alphaeff}{\pi} \right)^3 {C_A^2 \over 12} \left( -\frac{5}{3} - \frac{\pi^2}{6} + \frac{\pi^4}{20} -\zeta_3 \right) + {\mathcal O}(\alphaeff^4 )\right]  \theta^2 + {\mathcal O}(\theta^4 ) \,.\nonumber
\end{align}

The observed $n_{f}$-independence of  $\Omega(\alphaeff,x)$ leads us to conjecture 
that the latter quantity
is universal in gauge theories, i.e. independent of the specific particle content of the theory.
Assuming this conjecture leads to a number of non-trivial predictions.

First, let us recall the known value for $K$ in  ${\mathcal N}=4$ super Yang-Mills (in the  $\overline{\rm{DR}}$ scheme) \cite{Kotikov:2004er}, 
\begin{equation}
\begin{aligned}
K_{\rm {{\mathcal N}=4}}(\alpha_s) =& \, C_{F} \left[ \left(\frac{\alpha_s}{\pi}\right) -\frac{\pi^2}{12} C_{A}  \left(\frac{\alpha_s}{\pi}\right) ^2 \right. \\ 
& \left.  \quad\quad  + \frac{11}{720} \pi^4 C_{A}^2 \left(\frac{\alpha_s}{\pi}\right) ^3  + {\mathcal{O}}(\alpha_s^4) \right] \,.
\end{aligned}
\end{equation}
Plugging this formula and the result for $\Omega$ given in eq. (\ref{result_omega})
into eq. (\ref{defomega}) then gives the previously unknown three-loop result for the  cusp anomalous dimension for the Wilson loop operator of eq.  (\ref{W-bosonic}) in that theory,
\begin{align}
 \Gamma_{ {\mathcal N}=4}(\alpha_s,x) =& 
 \frac{\alpha_{s}}{\pi} \, C_{F} \tilde{A}_1   + \frac{C_{A} C_{F} }{2} \left( \frac{\alpha_{s}}{\pi}\right)^2 \, \left[ \tilde{A}_3+ \tilde{A}_2 \right] \nonumber  \\
& \hspace{-2.5 cm} +\frac{C_F C_A^2}{4} \left( \frac{\alpha_{s}}{\pi}\right)^3\,  \left[ \tilde{A}_5+ \tilde{A}_4 -\tilde{A}_2 + \tilde{B}_5 
+ \tilde{B}_3  \right] 
+ {\mathcal O}(\alpha_s^4) \,.
\end{align}
The two-loop terms agree with ref. \cite{Grozin:2014axa}.
As a test of the three-loop prediction, we take the antiparallel lines limit and obtain
\begin{equation}
\begin{aligned}
 \Gamma_{ {\mathcal N}=4}(\alpha_s,x)  \stackrel{\delta \to 0}{=}{} &  -\frac{C_{F} \alpha_s }{\delta} \left\{ 1 - \left(\frac{\alpha_s}{\pi}\right) C_{A}   \right.   \\
& \left.  
\hspace{-2.6cm} 
+ \left(\frac{\alpha_s}{\pi}\right)^2 C_{A}^2 \left[ \frac{5}{4}+ \frac{\pi^2}{4} - \frac{\pi^4}{64} \right] + {\mathcal{O}}(\alpha_s^3) \right\} + {\mathcal{O}(\delta^0)} \,,
\end{aligned}
\end{equation}
as expected from the direct calculation of the quark anti-quark potential \cite{Prausa:2013qva}. 

Second, the conjecture of the $n_f$-independence of $\Omega$ can be used to
predict the form of the non-planar $n_{f}$ corrections that can first appear at four loops.
The latter involve quartic Casimir operators of $SU(N)$, whose contribution we abbreviate by
$C_{4} = d^{abcd}_{F} d^{abcd}_{F}/N_{A} =  (18- 6 N^2 + N^4)/(96 N^2)$ (with $N_A$ the number of the $SU(N)$ generators) \cite{vanRitbergen:1997va}.
Consider a term in $\Gamma_{\rm cusp}(\alpha_s, x)$ of the form $n_f  (\alpha_s/\pi)^4 g(x) C_F C_{4}/64$, for some $g(x)$. Assuming that $\Omega$ defined in eq. (\ref{defomega}) is independent of $n_f$ then implies $g(x) =  g_0  \tilde{A}_{1}$.
Moreover, we can determine $g_{0}$ by comparing to the antiparallel lines limit.
The expected relation to the known quark antiquark potential computed (numerically) in ref. 
\cite{Smirnov:2008pn} then yields $g_{0} = -\, 56.83(1)$.

Finally, if $\Omega$ is the same in any gauge theory, 
it is likely that it will be determined in the foreseeable future, at least in the planar limit, by calculations based on integrability in ${\mathcal N}=4$ super Yang-Mills.

\medskip
{\it Acknowledgments}

A.G.'s work was supported by RFBR grant 12-02-00106-a and by the
Russian Ministry of Education and Science.  J.M.H. is supported in
part by the DOE grant DE-SC0009988, the Marvin L. Goldberger fund, and
the Munich Institute for Astro- and Particle Physics ${\rm (MIAPP)}$
of the DFG cluster of excellence ``Origin and Structure of the
Universe''.  G.P.K. is supported in part by the French National Agency
for Research (ANR) under contract StrongInt (BLANC-SIMI-4-2011).
P.M. was supported in part by the DFG through the SFB/TR 9
``Computational Particle Physics'' and the EU Networks LHCPHENOnet
PITN-GA-2010-264564 and HIGGSTOOLS PITN-GA-2012-316704.  We thank
J. Bl\"umlein  and J. Ablinger for making the ``HarmonicSums'' computer
code based on \cite{HarmonicSums} available to us.

\end{document}